\documentclass[10pt]{article}
\usepackage[cp1251]{inputenc}
\usepackage[russian]{babel}
\usepackage{graphicx}

\begin{document}

\twocolumn[\noindent{\small\it  ISSN 1063-7729, Astronomy Reports,
Vol. 50, No. 9, 2006, pp. 733--747. \copyright Pleiades
Publishing, Inc., 2006. \noindent Original Russian Text \copyright
V.V. Bobylev, G.A. Gontcharov, and A.T. Bajkova, 2006, published
in Astronomicheski$\check{\imath}$ Zhurnal, 2006, Vol. 83, No. 9,
pp. 821--836. }

\vskip -4mm

\begin{tabular}{llllllllllllllllllllllllllllllllllllllllllllllll}
 & & & & & & & & & & & & & & & & & & & & & & & & & & & & & & & & & & & & & & & \\
\hline \hline
\end{tabular}

\vskip 1.5cm

\centerline{\large\bf The OSACA Database and a Kinematic Analysis
of Stars}\centerline{\large\bf in the Solar Neighborhood }

\bigskip

\centerline{\bf  V. V. Bobylev, G. A. Gontcharov, and A. T.
Bajkova}

\medskip

\centerline{\it Main (Pulkovo) Astronomical Observatory, Russian
Academy of Sciences,} \centerline{\it Pulkovskoe sh. 65,
St.Petersburg, 196140 Russia} \centerline{\small Received November
22, 2005; in final form, April 14, 2006}

\vskip 1.5cm

{\bf Abstract} --- {\small We transformed radial velocities
compiled from more than 1400 published sources, including the
Geneva--Copenhagen survey of the solar neighborhood (CORAVEL-CfA),
into a uniform system based on the radial velocities of 854
standard stars in our list. This enabled us to calculate the
average weighted radial velocities for more than 25~000 HIPPARCOS
stars located in the local Galactic spiral arm (Orion arm) with a
median error of $\pm 1$ km s$^{-1}$. We use these radial
velocities together with the stars\' coordinates, parallaxes, and
proper motions to determine their Galactic coordinates and space
velocities. These quantities, along with other parameters of the
stars, are available from the continuously updated Orion Spiral
Arm CAtalogue (OSACA) and the associated database. We perform a
kinematic analysis of the stars by applying an Ogorodnikov-Milne
model to the OSACA data. The kinematics of the nearest single and
multiple main-sequence stars differ substantially. We used distant
($r \approx 0.2$ kpc) stars of mixed spectral composition to
estimate the angular velocity of the Galactic rotation, $\omega_o
= -25.7\pm 1.2$ km s$^{-1}$ kpc$^{-1}$, and the vertex deviation,
$l = 13^o \pm  2^o$, and detect a negative K effect. This negative
K effect is most conspicuous in the motion of A0--A5 giants, and
is equal to $K=-13.1 \pm 2.0$ km s$^{-1}$ kpc$^{-1}$. }

\bigskip


{\bf DOI}: 10.1134/S1063772906090071

\vskip 1cm

]

\centerline{1.~INTRODUCTION }

\medskip

Radial velocities of several tens of thousands of stars have been
measured in recent years with precisions of better than $\pm$ 1 km
s$^{-1}$, making joint analysis of stellar proper motions and
radial velocities of considerable importance for studies of
stellar kinematics. An appreciable fraction of these
high-precision radial velocities were measured using the CORAVEL
photoelectric cross-correlation spectrometers and digital
spectrometers of the Harvard-Smithsonian Center for Astrophysics
(CfA) and can be found in the catalog of Nordstr$\ddot{o}$m et al.
[1] (hereafter, the CORAVEL-CfA catalog). These authors and
Gontcharov [2] showed that the combined use of proper motions and
radial velocities, interstellar extinction estimates, multicolor
photometry, and binary parameters makes it possible to draw
conclusions about the absolute magnitudes, ages, masses, and
Galactic orbits of the stars and the evolution of Galactic
structures over millions of years, and to perform other
stellar-kinematic studies. In order to systematically collect,
analyze, and reduce the corresponding observational material, we
have created and have been continuously updating our database for
stars with known coordinates, parallaxes, proper motions, and
radial velocities.

Although the mutual alignment and density of spiral arms in the
solar neighborhood require a special study, intrinsically bright
HIPPARCOS stars, which are tracers of the Galactic spiral arms,
enable us to estimate fairly confidently the position and size of
the local spiral arm. The HIPPARCOS catalog is complete to about
$7.3^m$ in the band of the Milky Way, i.e., for stars with
absolute magnitudes $M < -2^m$ located out to 600 pc, given that,
as a rule, the extinction is lower than $0.5^m$ in this region.
Moreover, stars with absolute magnitudes $M < -4^m$ belonging to
neighboring spiral arms are represented almost completely in this
catalog. The distribution of intrinsically bright stars over the
Galactic octants and in heliocentric distance presented in Table 1
indicates that, first, nearby intrinsically bright stars are
distributed almost uniformly over the Galactic octants; second,
such stars are not distributed uniformly in Galactic longitude at
greater distances; and, third, the distribution changes abruptly
at a heliocentric distance of more than 700 pc. Thus within 400 pc
from the Sun, only the Scorpio.Sagittarius association can be
identified near a Galactic longitude of about 340.. At distances
between 400 to 700 pc, intrinsically bright stars are mostly
concentrated near latitudes of $90^o$ and $225^o$, possibly
indicating the directions to which  the local spiral arm extends,
as well as the width of the arm, which is about $\pm$500 pc in the
directions of the Galactic center and anticenter. The predominant
longitudes change beyond 800 pc: the regions of enhanced density
of intrinsically bright stars near $110^o$ and $290^o$ correspond
to the well-known Perseus and Sagittarius arms, respectively. We
can thus refer to the region within 500 pc of the Sun as the
portion of the local Orion arm that surrounds us.

Currently, radial velocities are known and parallaxes are
relatively accurate primarily for stars within 500 pc of the Sun,
i.e., within the local Galactic spiral arm (Orion arm). For this
reason, we called the catalog based on our database the Orion
Spiral Arm CAtalog.OSACA (http://www.geocities.com/orionspiral/).
The database, with its numerous tables, relations, and queries is
a sort of "factory", whereas the final product that can be
distributed is the OSACA catalog, which consists of several tables
that are periodically compiled from the database.

The goal of this paper is to describe the catalog and to analyze
the most interesting examples of its use for studies of the
kinematics of stars in the solar neighborhood. One such example is
analyzing the effect of an adopted model for the binarity of the
nearest stars on their kinematic parameters. As another example,
we considered relatively distant A0.A5 stars exhibiting a negative
K effect. We performed our kinematic analysis using a
Ogorodnikov-Milne model.We determined the Schwarzschild ellipsoid
to analyze the residual velocities of the stars and construct
two-dimensional distributions in the $U,V$ plane.

\vskip 1cm

\centerline{2.~CONTENT AND FORMAT}\centerline{OF THE OSACA
CATALOG}

\medskip

The core of the OSACA catalog are the six quantities that describe
the position and velocity of a star in space: $\alpha, \delta,
\pi, \mu_\alpha, \mu_\delta,$ and $V_r$. These quantities are used
to calculate the coordinates $X, Y, Z$ and space velocities $U, V,
W$ [3] in the standard Galactic Cartesian coordinate system, where
X increases toward the Galactic center, $Y$ in the direction of
Galactic rotation, and $Z$ in the direction of the North Galactic
Pole. The OSACA catalog is continuously updated with new published
data. Every substantial change in the input data is followed by a
recalculation of the dependent quantities; for example, the
velocity components $U, V,$ and $W$ are recomputed after a
refinement of the radial velocities in the database.

Since positions and velocities are known sufficiently precisely
only for stars of the HIPPARCOS catalog [4], the OSACA catalog
currently includes only these stars, although we collect data for
other stars in our database as well. The proper motions of the
HIPPARCOS stars considered have median errors in proper motion of
$\pm$0.0007$''$ yr$^{-1}$, in right ascension and declination of
$\pm$0.0006$''$ yr$^{-1}$, and in the tangential velocity of
$\pm$0.0009$''$ yr$^{-1}$. The proper motions of some of the stars
considered have low accuracy due to their binarity. To refine
these proper motions, we used a combination of coordinate and
proper-motion data from the HIPPARCOS catalog and from
ground-based catalogs, in accordance with the method of Gontcharov
and Kiyaeva [5]. The parallaxes, equatorial coordinates, and
proper motions of the stars considered are updated comparatively
infrequently, when new data whose accuracy are superior to the
HIPPARCOS data become available. Thus, the main data that are
collected and refined are the stellar radial velocities.

The catalog currently includes the main table, a table of
references to published radial velocities (one line per star), a
table of references to published radial velocities (one line per
publication), and a table of cross identifications of the stars.
Table 2 gives the format of the main table of the current release
of the catalog. Figure 1 shows the distribution of the stars in
absolute magnitude and heliocentric distance for single (solid)
and multiple (dashed) stars. The distribution of the stars in the
absolute magnitude. B.V color index diagram (Fig. 2) resembles the
corresponding distribution for all stars of the HIPPARCOS
catalog.all spectral types and luminosity classes are represented.

\vskip 1cm

\centerline{3.~CALCULATION OF THE RADIAL}\centerline{VELOCITIES}

\medskip

The WEB catalog [8] contains about 14 300 HIPPARCOS stars with
precise parallaxes and radial velocities known to better than
$\pm$5 km s$^{-1}$. The catalog of Barbier-Brossat and Figon [9]
contains about 16 000 such stars. Our catalog contains more than
28 000 such stars, reflecting the large number of high-precision
radial-velocity measurements obtained in recent years, including
the measurements for more than 10 000 stars reported in the
CORAVEL-CfA catalog, as well as our reduction of all the
radial-velocity catalogs employed to a single system.

We calculated the weighted mean radial velocities using data from
more than 1400 publications. The full list of these sources can be
found on the OSACA web page
(http://www.geocities.com/orionspiral/). Examples of extensive
catalogs that we employed include [10.21] and all the sources used
to produce the catalogs [8, 9, 22]. All the sources of the radial
velocities can be subdivided into two parts: (1) about one hundred
major source catalogs, each of which usually has its own system of
radial velocities, and (2) over one thousand publications
dedicated to individual stars or small groups of stars. In the
former case, the systematic errors for a particular catalog can be
revealed by comparing it to some standard or other catalog,
whereas such a comparison is impossible in the latter case due to
the small number of stars. If individual observations of stars
were made with the same instrument as the measurements reported in
some major catalog, we can hope that the radial-velocity system
will be the same for these observations, so that the systematic
corrections assigned to the instrument in question can be applied
to that individual star.

The problem of radial-velocity standards has a long history, and
has become especially topical in recent years (see, e.g.,
[23.27]). At present, only 181 stars approved by the International
Astronomical Union (IAU) as radial-velocity standards have been
confirmed to have constant radial velocities. The HD and HIPPARCOS
numbers, radial velocities, V magnitudes and absolute magnitudes,
and B.V color indices of these stars are given in the table posted
on the OSACA web page; in these cases, the "source" column
indicates "IAU".

However, the need to use certain stars as radial-velocity
standards has been questioned in recent years due to the emergence
of astrometric methods of measuring radial velocities [16] and the
fundamental possibility of calibrating an instrument by observing
the radial velocities of Solar-system bodies.

Nidever et al. [27] (which we will refer to as the Keck catalog)
shows that: (1) the use of standard stars and Solar-system bodies
yields virtually the same radial velocities to within $\pm$0.5 km
s$^{-1}$, (2) it is possible to create a list of several hundred
standard stars for determining radial velocities with errors no
greater than $\pm$0.5 km s$^{-1}$, which would reasonably
represent the entire variety of spectral types, luminosity
classes, and other stellar parameters, and (3) determining the
absolute radial velocities to better than $\pm$0.5 km s$^{-1}$
requires allowance for numerous effects that thus far lack a
sufficient theoretical basis (to say nothing of the procedures for
taking them into account in practice!).

Stellar kinematic studies require not so much radial velocities
with very high precision as a homogeneous data set of radial
velocities for a large number of various stars. A precision of 1
km s$^{-1}$ can be considered to be acceptable for such studies. A
velocity of 1 km s$^{-1}$ corresponds to 1 pc Myr$^{-1}$. This
precision enables the calculation of Galactic orbits and the
analysis of the evolution of Galactic structures and groups of
stars within 500 pc of the Sun over several tens of million of
years in the past and future. Moreover, within a large volume of
space, a precision of 1 km s$^{-1}$ is comparable to the median
accuracy of stellar tangential velocities calculated from proper
motions, namely, 0.001 $''$ yr$^{-1}$; within 200 pc of the Sun,
the tangential velocities are, on average, known better than the
corresponding radial velocities (for about two-thirds of all the
stars considered), whereas the situation is reversed for more
distant stars (about one-third of all the stars).

The Keck catalog includes a list of 782 stars proposed for use as
standard stars for determinations of radial velocities. These
stars have stable relative radial velocities, and represent
different types of stars; the radial velocities of IAU standard
stars in the Keck catalog agree with the standard radial
velocities to within no worse than $\pm$0.5 km s$^{-1}$. Although
some of these stars exhibited radial-velocity variations after the
release of the Keck catalog, the overwhelming majority can be
adopted as supplementary standards for the determination of radial
velocities with a precision of $\pm$(0.5-1) km s$^{-1}$. The
parameters of 673 such stars with "Keck" indicated in the "source"
column are given in the table posted on the OSACA web page.

The full list of 854 standard stars that we used to transform more
than 50 000 published radial velocities to a single standard
system that we refer to as the "IAU--Keck" system does not include
some stars in the initial IAU and Keck lists that have exhibited
appreciable radial-velocity variations in recent years. Figure 3
shows the distribution of these stars on the $(B-V )-M_V$ color
index.absolute magnitude diagram. It is evident that types of
stars that are not represented or under-represented include white
dwarfs, subdwarfs, subgiants, and A5 to F5 main-sequence stars.

The transformation of a published radial-velocity catalog into the
standard system involves the calculation of "catalog minus
standard" differences for stars in common followed by the
identification and approximation of systematic dependences of
these differences on various stellar parameters allowing the
systematic differences identified to be eliminated for all stars.

As an example, let us consider the systematic errors that we
revealed for the CORAVEL-CfA catalog. Figure 4 shows the
dependences of the radial-velocity differences in the sense
"CORAVEL-CfA minus standard" for 484 stars in common as a function
of radial velocity, B.V color index, and absolute magnitude (we
also found the dependences on effective temperature, $H_{\beta}$
index, and metallicity Fe/H). The left- and right-hand plots show
the data before and after elimination of the following
dependences:

$$\Delta V_{r}=-0.8398*(B-V)+0.2531,$$
$$\Delta V_{r}=0.000008*V_{r}^{2}+0.0012*V_{r}-0.0142,$$
where $\delta V_r$ is the difference of the radial velocities for
stars in common in the sense CORAVEL-CfA minus standard, B.V is
the color index, and $V_r$ is the radial velocity. It is obvious
from these plots that eliminating the dependences on radial
velocity and $B-V$ color index also eliminated the dependence on
absolute magnitude (and on other parameters as well); no new
dependences appeared in the process, and the standard deviation of
the CORAVEL-CfA minus standard radial-velocity differences
decreased from 0.38 to 0.33 km s$^{-1}$ for the stars in common.
This quantity can be viewed as an approximate estimate of the
systematic accuracy of the CORAVEL-CfA catalog.

We used the same procedure to bring other catalogs to the IAU.Keck
standard system. As for the stars of the CORAVEL-CfA catalog, the
most widespread systematic errors for these stars depend on radial
velocity and $B-V$. Dependences on these two parameters are
usually manifest in observations of radial velocities (see, e.g.,
[27]).

After transforming all the radial velocities to the single system,
we calculate for each star its weighted mean radial velocity. In
this case, we set the weights in accordance with the standard
deviation of the catalog minus standard radial-velocity
differences for stars in common to the catalog and standard. We
assign single observations the same weight as that for catalogs
obtained from observations performed with the same instrument.

The radial velocities of the CORAVEL-CfA catalog transformed to
the standard IAU.Keck system have such high precisions compared to
many other sources and the CORAVEL-CfA catalog has such high
weights that the weighted mean velocities for these stars barely
differ from the corresponding CORAVEL-CfA radial velocities with
the systematic errors corrected.

However, some of the CORAVEL-CfA data should be treated with
caution. Figure 5 compares the radial velocities and $U, V, W$
space-velocity components from the CORAVEL-CfA catalog and from
our calculations based on other two sources of radial velocities
for 3400 stars in common (all radial velocities were transformed
to a single system and the same HIPPARCOS/Tycho-2 proper motions
were used). All the quantities are appreciably correlated. The
differences in the radial velocities are due mostly to errors in
catalogs that are less accurate than the CORAVEL-CfA catalog.
However, a fairly large number of stars exhibit discrepances
between the  $U, V$, and $W$ values from the CORAVEL-CfA catalog
and our calculations. This discrepancy is due to the distances
employed: we used distances calculated from the HIPPARCOS
trigonometric parallaxes, whereas photometric parallaxes were used
to calculate the distances in the CORAVEL-CfA catalog and the
corresponding factor for transforming the proper motions into
linear velocities for stars for which the relative errors of the
HIPPARCOS parallaxes is greater than 13\%. The authors of the
CORAVEL-CfA catalog admit that these photometric parallaxes are
not suitable for binary stars, giants, and several other
categories of stars. We found most of the stars with fairly
discordant $U, V$, and $W$ velocity components for the CORAVEL-CfA
catalog and our own data, indeed, to be such "problematic"
objects, whose photometric parallaxes in the CORAVEL-CfA catalog
are systematically greater than the corresponding trigonometric
parallaxes (and whose heliocentric distances are systematically
smaller). As is evident from Fig. 6, the same effect arises for
the same "problem" stars when we compare the distances based on
the HIPPARCOS parallaxes and the CORAVEL-CfA distances, and in the
appreciable difference between our absolute magnitudes and the
absolute magnitudes adopted from the CORAVEL-CfA catalog (note
that the discrepancy between the interstellar extinctions inferred
using different methods is insignificant, since we are dealing
with fairly nearby stars whose extinction does not, on average,
exceed 0.1$^m$).

Among the stars considered here are more than 7000 known or
suspected multiple stars, including more than 800 known and 4500
suspected spectral binaries. Therefore, one of the problems we had
to face when calculating the weighted mean radial velocities were
the errors in the identification of components of multiple systems
made in many publications. In other words, authors often
incorrectly identify the  component whose radial velocity they
have measured, or even fail to mention the component to which the
measurement refers. These errors are largely due to the fact that,
until recently, the nomenclature for stellar and planetary systems
was not standardized; i.e., the same component had different
designations in different catalogs. We made a number of
corrections of this type.

In all cases where there is no doubt that the components of a
stellar system are gravitationally bound (the component separation
does not exceed 20 000 AU), we calculated only the systemic radial
velocities, and will consider only these velocities further. In
the case of gravitationally wide stellar pairs (with component
separations exceeding 20 000 AU), the OSACA catalog lists all
components with measured radial velocities.

Estimates of the errors in the radial velocities are available for
most of the stars in the OSACA database, enabling us to estimate
the random errors in the observed space velocities. For example,
26 492 stars satisfy the criteria

$$
\displaylines{\hfill
 e_\pi/\pi<0.3,              \hfill\llap{(1)}\cr\hfill
     e_{V}< 15~\hbox{km s$^{-1}$},  \hfill\llap{(2)}\cr\hfill
   V_{pec}<100~\hbox{km s$^{-1}$},  \hfill\llap{(3)}
   }
$$
Here, $e_\pi$ is the fractional error in the parallax, $e_V =
\sqrt{e^2_{V_r} + e^2_{V_l} + e^2_{V_b}}$ is the error in the
observed velocity vector $V = (Vr, Vl, Vb)$, where $V_r$ is the
radial component of the velocity, $V_l = 4.74\mu_l \cos br$ is the
velocity component in the direction of Galactic longitude, and
$V_b = 4.74\mu_b r$ is the velocity component in the direction of
Galactic latitude; $V_{pec}=\sqrt{U^2+V^2+W^2}$ is the residual
(peculiar) velocity of the star relative to the local standard of
rest. We estimated the errors in $V_l$ and $V_b$ taking into
account the errors in the parallax but without allowance for
correlations between the proper-motion components $\mu_l\cos b$
and $\mu_b$, using the formula

$$
\displaylines{\hfill
 e_{(V_l, V_b)} = {4.74\over\pi}
 \sqrt{\mu^2_{l,b}\cdot\Biggl({e_\pi\over\pi}\Biggr)^2+e^2_{\mu_{l,b}}},\hfill
 }
$$
where $\mu_{l,b}$ are the tangential-velocity components. We set
the velocity of the Sun relative to the local standard of rest
equal to $(u_\odot,v_\odot,w_\odot)=(10.0,5.3,7.2)$ km s$^{-1}$,
in accordance with [28]. We found the mean random errors to be
${\overline e_{V_r}}=2.7$~km s$^{-1}$,
 ${\overline e_{V_l}}=2.3$~km s$^{-1}$, and
 ${\overline e_{V_b}}=1.7$~km s$^{-1}$.

\vskip 1cm

\centerline{4.~METHODS FOR THE KINEMATIC}\centerline{ANALYSIS}

\medskip

In this paper, we analyze the kinematics of stars based on the
OSACA catalog in an Ogorodnikov-Milne model.We analyze the
residual velocities using the classic method of determining the
Schwarzschild ellipsoid and two-dimensional distributions of the
$U, V$ velocity components.

\medskip

\centerline{\it 4.1.~The Ogorodnikov-Milne Model}

\medskip

We used a linear Ogorodnikov-Milne model with the notation
introduced by Clube [29, 30]. The observed velocity ${\bf V}(r)$
of a star with a heliocentric radius vector ${\bf r}$ can be
described by the following vector equation, up to first-order in
the small quantity $r/R \ll 1$:

$$
\displaylines{ \hfill
 {\bf V}(r) ={\bf u}_\circ+M{\bf r}+{\bf V'},\hfill\llap
 }
$$
where ${\bf u}_\circ$ is the velocity of the centroid of the stars
relative to the Sun ($-V_\odot(u_\odot,v_\odot,w_\odot)$), {$\bf
V'$} is the residual velocity of the star (here we assume that the
stellar residual velocities are randomly distributed), and $M$ is
the matrix (tensor) of the displacements, whose components are
equal to the partial derivatives of ${\bf u}(u_1,u_2,u_3)$ with
respect to ${\bf r}(r_1, r_2, r_3)$:
$$
\displaylines{\hfill
 M_{pq}={\left(\frac{\partial u_p}{\partial r_q}\right)}_\circ,
 \quad (p,q=1,2,3). \hfill\llap
}
$$
The matrix M can be decomposed into the local deformation tensor
$M^+$ and the local-rotation tensor $M^-$ [31]:
$$
\displaylines{\hfill
 M_{\scriptstyle pq}^{\scriptscriptstyle+}=
   {1\over 2}\left( \frac{\partial u_{p}}{\partial r_{q}}+
    \frac{\partial u_{q}}{\partial r_{p}}\right)_\circ,\hfill\llap(4)\cr\hfill
  M_{\scriptstyle pq}^{\scriptscriptstyle-}=
  {1\over 2}\left(\frac{\partial u_{p}}{\partial r_{q}}-
   \frac{\partial u_{q}}{\partial r_{p}}\right)_\circ,~~(p,q=1,2,3).\hfill\llap
     }
$$
The working equations have the form
$$
\displaylines{\hfill
 V_r= f(u_\odot,v_\odot,w_\odot,M_{11},...,M_{33}),\hfill\llap(5)\cr\hfill
 V_l= f(u_\odot,v_\odot,M_{11},...,M_{23}),        \hfill\llap(6)\cr\hfill
 V_b= f(u_\odot,v_\odot,w_\odot,M_{11},...,M_{33}),\hfill\llap(7)
 }
$$
(see [32] for more details). The set of conditional equations
(5).(7) contains 12 unknowns: three velocity components for
$V_\odot$ and nine components of $M_{pg}$, which are estimated
using the least-squares method. We then calculate the components
of the deformation and rotation tensors using the already derived
$M_{pg}$ values based on (4). Thus, the Oort constants can be
calculated as follows:  $A=0.5(M_{12}+M_{21})$,
 $B=0.5(M_{21}-M_{12})$,
 $C=0.5(M_{11}-M_{22})$ и
 $K=0.5(M_{11}+M_{22})$, and the vertex deviation $l_{xy}$
can be calculated from the formula  $\tan 2 l_{xy}=-C/A$. The
angular velocity of the Galactic rotation can be obtained from the
formulas $\omega_\circ=B-A=-M_{12}$. The form of these equations
is such that negative $\omega_\circ$ corresponds to the direction
of the Galactic rotation.

\medskip

\centerline{\it 4.2.~Determination of the Schwarzschild Ellipsoid}

\medskip

We calculated the elements of the residual-velocity ellipsoid
(Schwarzschild ellipsoid) using the well-known statistical method
described in detail in [31, 33, 34]. This method consists in
determining the symmetric momentum tensor. We have six equations
for each star that can be used to determine the six unknown
components of the velocity-dispersion tensor. Analysis of the
eigenvalues of the velocity dispersion tensor yields the principal
semiaxes $\sigma_{1,2,3}$ of the residual-velocity ellipsoid
$\sigma_{1,2,3}$ and also the directions $l_{1,2,3}$, $b_{1,2,3}$
of the principal axes. When writing the formulas for the stellar
residual velocities, we take into account the overall rotation of
the Galaxy by adopting the Oort constants $A = 13.7 $km s$^{-1}$
kpc$^{-1}$ and $B = -12.9$ km s$^{-1}$ kpc$^{-1}$, in accordance
with [35].

\medskip

\centerline{\it 4.3.~Two-dimensional Maps of the Space Velocities}

\medskip

It is of interest to construct and analyze two-dimensional
velocity maps, in particular, velocity maps in the $U,V$ plane
[36, 37], in order to analyze the fine structure of the
distributions of the residual space velocities in order to
identify clusters, streams, etc. We estimated the two-dimensional
probability density $f(U,V)$ from the calculated, discretely
distributed U, V velocities using the method of adaptive smoothing
[36] with a two-dimensional radially symmetric Gaussian kernel:
$$
K(r)=\frac{1}{\sqrt{2\pi}\sigma} \exp{-{\frac{r^2}{\sigma^2}}},
$$
where $r^2=x^2+y^2$. In this case, the necessary relation $\int
K(r)dr=1$ for probability density estimates is satisfied. The main
idea behind this method is that, at each point of the map,
smoothing is performed by a beam whose size is determined by a
parameter $\sigma$, which varies depending on the data in the
neighborhood of the point considered. Thus, the smoothing is
performed using a fairly narrow beam in high-density regions, and
the beam width increases with decreasing density. We use the
following definition for adaptive smoothing at an arbitrary point
$\xi=(U,V)$ [36]:
$$
 \hat{f}(\xi)=\frac{1}{n}\sum_{i=1}^n K\left(\frac{\xi-\xi_i}{h\lambda_i}\right),
$$
where $\xi_i=(U_i,V_i)$, $\lambda_i$ is a local dimensionless
parameter of the beam at the point $\xi_i$, $h$ is a general
smoothing parameter, and $n$  is the number of data points
$\xi_i=(U_i,V_i)$. The parameter $\lambda_i$ at each point of the
two-dimensional $U,V$ plane is defined as follows:
$$
 \lambda_i=\sqrt\frac{g}{\hat{f}(\xi)}~,~~~~~~~
 \ln g=\frac{1}{n}\sum_{i=1}^n \ln \hat{f}(\xi),
$$
where $g$ is the geometric mean $\hat{f}(\xi)$. It is evident
that, to determine $\lambda_i$, we must know the distribution
$\hat{f}(\xi)$, which, in turn, can be determined if all
$\lambda_i$ are known. Therefore, the problem of finding the
unknown distribution must be solved iteratively. As a first
approximation, we use the distribution obtained by smoothing the
initial $U,V$ map with a fixed-sized beam. The optimum value of
the parameter $h$ can be determined by minimizing the mean square
deviation of the estimate $\hat{f}(\xi)$ from the true
distribution $f(\xi)$. The h values that we inferred for the
various stellar samples considered here vary from 6 to 10 km
s$^{-1}$. A typical uncertainty in the velocity is 2 km s$^{-1}$
in this case. This influenced our choice of the discretization
interval for the two-dimensional maps; the area of a square pixel
was taken to be $S=2\times 2$  km$^2$ s$^{-2}$. The map must be
scaled using the factor $nS$.

\vskip 1cm

\centerline{5.~RESULTS OF THE KINEMATIC ANALYSIS}

\medskip

\centerline{\it 5.1.~Effect of Binary Stars}

\medskip

The aim of this section is to establish how the binarity of stars
affects the parameters of the Ogorodnikov-Milne model. To identify
possible features that depend on stellar age, we selected from the
OSACA catalog stars of luminosity classes V and IV, i.e.,
main-sequence stars satisfying the conditions
$$
 e_\pi/\pi<0.1, ~~~~~ e_{V}< 8~\hbox{km s$^{-1}$},
$$
and also condition (3). Figure 7 shows the distribution of random
errors of the $V_r,V_l,V_b$ components of the observed space
velocities as a function of distance for the resulting sample of
6276 stars. We find the mean random errors of the velocity
components to be ${\overline e_{V_r}}=1.1$ km s$^{-1}$,
${\overline e_{V_l}}=1.3$ km s$^{-1}$, ${\overline e_{V_b}}=0.9$
km s$^{-1}$. In this case, the mean error in the absolute space
velocity is 1.9 km s$^{-1}$. The homogeneous nature of our data
enables us to solve the initial equations (5)--(7) with unit
weights.

We first subdivided the entire sample into two parts depending on
the color index, with the boundary at $B-V = 0.5$, then produced
for each part subsamples consisting of either single or binary
stars. Table 3 gives our derived components of the solar velocity,
$V_\odot(u_\odot,v_\odot,w_\odot)$ relative to the corresponding
stellar centroid, coordinates $L_\odot$ and $B_\odot$ of the apex
of the solar motion, elements $M_{pq}$ of the displacement tensor
and roots $\lambda_{1,2,3}$ of the deformation tensor, principal
axes $\sigma_{1,2,3}$ of the residual-velocity ellipsoid, and
directions $l_{1,2,3}, b_{1,2,3}$ of the principal axes. As is
evident from the upper part of the table, the components of the
solar motion based on the samples of single and binary stars do
not differ significantly in any of the color intervals.

The only elements of the displacement tensor to differ
significantly from zero are $M_{12}$ for the three samples
indicated in columns 2--4 and $M_{11}$ for the last sample of
binary main-sequence red dwarfs (column 5). The $M_{11}$ values
derived for the samples of single and binary red dwarfs differ by
$\Delta M_{11}\cdot{\overline r}\approx3\pm1$ km s$^{-1}$.

Two roots of the deformation tensor, $\lambda_{1,2}$, differ
significantly from zero for the two single-star samples (columns 2
and 4), whereas the third root, $\lambda_3$, does not differ
significantly from zero. Hence, we have a planar case where
$\lambda_{(1,2)}=K\pm\sqrt{A^2+C^2}$ [38]. The closeness of the
absolute values of the roots $\lambda_1$ and $\lambda_2$ is
suggestive of rotation (without expansion/ contraction).

The deformation tensor for the binary red dwarfs represents a
linear case with a single significant root of $\lambda_1=117\pm29$
km s$^{-1}$ kpc$^{-1}$, and with the fairly high value $K=55\pm19$
km s$^{-1}$ kpc$^{-1}$. In this case, the ellipse of the
deformation tensor is highly elongated along the $x$ axis, i.e.,
we have expansion along a single axis.

As is evident from Table 3, the elements of the residual-velocity
ellipsoids derived for the single and binary samples do not differ
significantly in any of the color intervals. The elements of the
residual-velocity ellipsoid derived for the binary stars have
larger random errors.

The samples considered are most refined in terms of the errors of
the heliocentric distances and velocity components of the stars.

Our distributions of $U,V$ (Fig. 8) show that the binary stars are
much more concentrated than the single stars (by about a factor of
two) near the local peaks associated with the Hyades and Pleiades
clusters. Note that the bottom contour and contour step in the
figures shown are 10\% of the maximum value.

The specific features of the parameters of the Ogorodnikov-Milne
model found by analyzing the samples of single and binary red
dwarfs, such as the statistical significance of $M_{11}$, are
consistent with our results based on an analysis of a
significantly smaller sample of nearby visual binaries [39]. Note
that we have determined the kinematic parameters with much higher
precision in this current work. The distributions of the $U,V$
velocities of both the single and multiple stars (Fig. 8), and of
the red dwarfs (the oldest stars), in particular, show that star
formation in the Galactic disk is influenced by some persistent
factor, which could be associated with the bar at the Galactic
center [40,41] and/or the spiral structure of the Galaxy [42].

\medskip

\centerline{\it 5.2.~Negative $K$ effect}

\medskip

The aim of this section is to analyze stars exhibiting the
strongest negative K effect. To this end, we selected only single
stars that obey relations (1)--(3) and the criterion $\pi < 10$
milliarcseconds, which leaves us mostly with giants. All the
selected stars are located at distances of 0.1--0.6 kpc. Our
analysis of 9468 stars ($\overline r=0.175$ kpc) of mixed spectral
composition yielded $A = 12.3 \pm 1.0 $km s$^{-1}$ kpc $^{-1}$, $B
= -13.4 \pm 1.0$ km s$^{-1}$ kpc $^{-1}$, $C = -6.1 \pm 1.0$ km
s$^{-1}$ kpc$^{-1}$, and $K=-4.1 \pm 1.0$ km s$^{-1}$ kpc$^{-1}$.
We found the vertex deviation to be $l = 13^o\pm 2^o$ and the
angular velocity of the Galactic rotation to be $\omega_o = -25.7
\pm 1.2$ km s$^{-1}$ kpc$^{-1}$. The derived values of the Oort
constants $A,B,$ and $\omega_o$ are, on the whole, fairly
consistent with the Galactic rotation parameters obtained in
various other studies based on much more distant stars than those
used here [35, 43, 44]. At the same time, we found $K$ to differ
significantly from zero, which is of considerable interest. A
negative $K$ effect in the Ogorodnikov-Milne model means that the
stellar system considered is in a state of contraction.

Several authors [35, 45] have found negative $K$ effects of
-(1--6) km s$^{-1}$ kpc$^{-1}$ in the motion of OB stars (out to
heliocentric distances of $\approx 4$ kpc). The nature of this
result is not entirely clear. Various hypotheses have been
suggested: it could be an effect due to the specifics of the
methods used to measure the stellar radial velocities [46], or a
result of the influence of the bar in the center of the Galaxy
[46], or the effect of the spiral structure [47,48].
Fern$\acute{a}$ndez et al. [45] tried to eliminate the negative K
effect by taking into account the spiral structure, but their
attempt failed, and the negative K effect remained.

Note that the youngest and nearest OB stars exhibit a
characteristic positive K effect [32, 49], and therefore we do not
consider these stars here.

In this work, we succeeded in forming a sample of stars with a
maximum negative K effect, which consists of 1269 stars with color
indices $B-V \leq 0.2$. We did not use stars of luminosity classes
V and IV or O and B stars, so that the sample in question contains
only stars of spectral types A0.A5 and luminosity classes I, II,
III. However, the bulk of the sample is made up of stars with no
luminosity classes given in the OSACA catalog. We then analyzed
the residual velocities of the stars corrected for the overall
Galactic rotation (see Section 3.2). Our analysis yielded for the
peculiar velocity of the Sun $u_\odot= 10.2\pm0.4$ km s$^{-1}$,
$v_\odot= 10.9\pm0.4$ km s$^{-1}$, $w_\odot=  6.6\pm0.4$ km
s$^{-1}$, $V_\odot= 16.4\pm0.4$ km s$^{-1}$, $L_\odot= 47\pm
2^\circ$, $B_\odot= 24\pm 2^\circ$, and for the components $M^+$
and $M^-$ of the tensor $M$ (in units of km s$^{-1}$ kpc$^{-1}$)
characterizing the residual kinematic effects
$$
\displaylines{\hfill M=\pmatrix
 { -28.0_{(3.2)}& -1.4_{(2.5)}& -0.2_{(4.1)}\cr
   -11.4_{(3.2)}&  1.9_{(2.5)}&  2.6_{(4.1)}\cr
     0.3_{(3.2)}&  0.2_{(2.5)}& -3.3_{(4.1)}\cr},\hfill\cr\hfill
M^{\scriptscriptstyle+}=\pmatrix
 {-28.0_{(3.2)}& -6.4_{(2.0)}&  0.0_{(2.6)}\cr
   -6.4_{(2.0)}&  1.9_{(2.5)}&  1.4_{(2.4)}\cr
    0.0_{(2.6)}&  1.4_{(2.4)}& -3.3_{(4.1)}\cr},\hfill 
    \cr\hfill
M^{\scriptscriptstyle-}=\pmatrix
  {           0&  5.0_{(2.0)}& -0.2_{(2.6)}\cr
   -5.0_{(2.0)}&            0&  1.2_{(2.4)}\cr
    0.2_{(2.6)}& -1.2_{(2.4)}&            0\cr}.\hfill}
$$
The roots of the deformation tensor
($\lambda_1,\lambda_2,\lambda_3$) are (3.5,-29.3,-3.5) km s$^{-1}$
kpc$^{-1}$, with standard errors of -3 km s$^{-1}$ kpc$^{-1}$. The
directions of the principal axes of the deformation tensor are
$$
\displaylines{ \hfill
 L_1= 282\pm2^\circ, \quad B_1=-12\pm1^\circ,\hfill\cr\hfill
 L_2=  12\pm4^\circ, \quad B_2= -1\pm26^\circ,\hfill\cr\hfill
 L_3= 285\pm4^\circ, \quad B_3= 79\pm6^\circ.\hfill}
$$
The vertex deviation in the $xy$ plane is $l_{xy}=-33\pm4^\circ$;
we also derived the parameters of the $K$ effect:
$K=0.5(M_{11}+M_{22})=-13.1\pm2.0$ km s$^{-1}$ kpc$^{-1}$, and
$C=0.5(M_{11}-M_{22})=-14.9\pm2.0$ km s$^{-1}$ kpc$^{-1}$.

The rotation tensor, $M^-$, contains no elements that differ
significantly from zero, leading us to conclude that the radial
component is the dominant effect in the residual motions of the
stars considered. In this case, we set one of the roots
($\lambda_3$) equal to zero and consider only the $xy$ plane.
Since ($(\partial V_R/\partial R)_{R_\circ}=\lambda_1$, we have
$(V_R/R)_{R_\circ}=\lambda_2$, where $R$ is the unknown distance
from the kinematic center. However, the relations
$$
\displaylines{\hfill
 K+C=  4\pm3~\hbox{km s$^{-1}$ kpc$^{-1}$},\hfill\cr\hfill
 K-C=-29\pm3~\hbox{km s$^{-1}$ kpc$^{-1}$},\hfill
}
$$
yield an estimate of the characteristic residual velocity of the
stars, $(K-C)\cdot {\overline r}=-5.4\pm0.6$ km s$^{-1}$, and also
$K=-12.9$ km s$^{-1}$ kpc$^{-1}$ and $C=16.4$ km s$^{-1}$
kpc$^{-1}$ (the sign has changed), which characterizes the radial
component of the residual stellar velocities in the case
considered. The resulting estimate, $-5.4\pm0.6$ km s$^{-1}$, is
consistent with the amplitude of the periodic radial oscillations
of the stellar residual velocities in the interarm space,
$f_R=7\pm1$ 1 km s$^{-1}$, found by Mel$'$nik et al. [50] based on
an analysis of the motions of OB associations.

The direction of the third axis, $B_3= 79\pm6^\circ$, indicates
compression in the Galactic plane. In fact, both roots of the
deformation tensor ($\lambda_1$ and $\lambda_2$) are close to
zero. This means that the compression occurs along a single axis
in the direction $12-192^\circ$, with its error being
$2^\circ$--$4^\circ$.

The principal axes of the residual-velocity tensor
($\sigma_1,\sigma_2,\sigma_3$) are ($16.09\pm0.56, 10.04\pm0.53,
6.89\pm0.47)$) km s$^{-1}$, and the directions of the principal
axes have the coordinates
$$
\displaylines{\hfill
 l_1=  27\pm4^\circ, \quad b_1=-1\pm0^\circ,\hfill\cr\hfill
 l_2= 117\pm6^\circ, \quad b_2=-1\pm2^\circ,\hfill\cr\hfill
 l_3=  80\pm6^\circ, \quad b_3=88\pm3^\circ.\hfill}
$$
We can see that the angles $l_1$, $l_{xy}$ and $L_2$ differ from
each other, which further emphasizes the fact that the effect of
the radial component is significant in the kinematics of these
stars.

Figure 8c shows the distribution of the residual $U,V$ velocities
for the stars considered. The stars are concentrated in only two
peaks.the so-called Sirius and Pleiades peaks. It is striking that
the velocity of the stars displaying a negative $K$ effect
relative to the local standard of rest ($v_\odot$) is fairly
small.

Rybka$'$s [51] analysis of the proper motions of about 40 000
giants of luminosity class III yielded $K=-7.5\pm1.8$ km s$^{-1}$
kpc$^{-1}$ for G5--K0 stars located beyond 250 pc (with their mean
distance being about 364 pc). Our result agrees well with that of
Rybka [51], leading us to conclude that the use of the
high-precision space velocities of the OSACA stars has enabled us
to detect this negative $K$ effect at much closer distances. Our
approach has also shown that we actually have compression along a
single axis, enabling us to come closer to understanding of the
nature of this effect.

\vskip 1cm

\centerline{6.~CONCLUSIONS}

\medskip

We have created and have been continuously updating the OSACA
(Orion Spiral Arm CAtalogue) database for stars with known
coordinates, parallaxes, proper motions, and radial velocities
located in the Orion arm of our Galaxy. We transformed radial
velocities adopted from over 1400 sources, including the
Geneva-Copenhagen survey of the solar neighborhood (CORAVEL-CfA),
into a single system of radial velocities based on 854 standard
stars from our list. This enabled us to calculate the weighted
mean radial velocities of more than 25 000 stars of the HIPPARCOS
catalog with a median error of 1 km s$^{-1}$. We use the
elimination of systematic errors of the CORAVEL-CfA catalog as an
example to demonstrate the efficiency of the proposed standard of
stellar radial velocities.

We have corrected numerous errors in the identification of
components of multiple systems made in many publications. We have
used the OSACA data to analyze the kinematics of 6276 nearby
main-sequence stars with space-velocity errors $\pm$2 km s$^{-1}$.
The use of an Ogorodnikov-Milne model and analysis of the
distribution of the $U,V$ velocities the of stars revealed
significant differences in the kinematics of single and multiple
stars. In particular, multiple stars are much more concentrated
than single stars in the vicinity of the peak associated with the
Hyades-Pleiades stream in the $U,V$ velocity plane.

Based on a sample of 9468 relatively distant
(($r~\approx~0.2$~kpc) stars of various spectral types, we
estimated the angular velocity of the Galactic rotation to be
$\omega_\circ=-25.7\pm1.2$ km s$^{-1}$ kpc$^{-1}$ (we found the
vertex deviation to be $l=13\pm2^\circ$). The radial component of
this velocity indicates a K effect, which shows up most
conspicuously in the motions of A0-A5 giants. We found for the
parameters of the Ogorodnikov-Milne model $K=-13.1\pm2.0$~km
s$^{-1}$ kpc$^{-1}$ and $C=14.9\pm2.0$~km s$^{-1}$ kpc$^{-1}$,
which we used to estimate the characteristic residual velocity of
the stellar sample considered, $-5.4\pm0.6$~ km s$^{-1}$. The
maximum compression is in the direction $12-192^\circ$, with an
error of $2^\circ-4^\circ$.

\vskip 1cm

\centerline{7.~ACKNOWLEDGMENTS}

\medskip

We are grateful to the referee for valuable comments, which
contributed to the improvement of the paper. This work was
supported by the Russian Foundation for Basic Research (project
code 05-02- 17047).

\vskip 1cm

\vskip 1cm

\centerline{REFERENCES}

\medskip

1. B. Nordstr$\ddot{o}$m, M. Mayor, J. Andersen, et al., Astron.
Astrophys. 418, 989 (2004).

2. G . A. Gontcharov, Order and Chaos in Stellar and Planetary
Systems, Ed. by G . G . Byrd, K. V. Kholshevnikov, A. A.
Myll$\ddot{a}$ri, et al., Astron. Soc. Pac. Conf. Ser. 316, 276
((2004)).

3. P. G.Kulikovski$\breve{i}$, Zvezdnaya astronomiya (Stellar
Astronomy) (Nauka, Moscow, 1985) [in Russian].

4. The HIPPARCOS and Tycho Catalogues, ESA SP-1200 (1997).

5. G .A. Gontcharov and O. V. Kiyaeva, Pis$'$ma Astron. Zh. 28,
302 (2002) [Astron. Lett. 28, 261 (2002)].

6. F. Arenou, M. Grenon, and A. Gomez, Astron. Astrophys. 258, 104
(1992).

7. C. Turon, M. Cr$\acute{e}$z$\acute{e}$, D. Egret, et al., The
Hipparcos Input Catalogue (ESA Publ. Div., Noordwijk, 1992), Vols.
1.7.

8. M. Duflot, P. Figon, and N. Meysonnier, Astron. Astrophys.,
Suppl. Ser. 114, 269 (1995).

9. M. Barbier-Brossat and P. Figon, Astron. Astrophys., Suppl.
Ser. 142, 217 (2000).

10. D. Bersier, G. Burki, M.Mayor, et al., Astron. Astrophys.,
Suppl. Ser. 108, 25 (1994).

11. C. Fehrenbach, M.Duflot, C.Mannone, et al., Astron.
Astrophys., Suppl. Ser. 124, 255 (1997).

12. J. Fernley and T. G. Barnes, Astron. Astrophys., Suppl. Ser.
125, 313 (1997).

13. C. Flynn and K. C. Freeman, Astron. Astrophys., Suppl. Ser.
97, 835 (1993).

14. S. Grenier, M. O. Baylac, L. Rolland, et al., Astron.
Astrophys., Suppl. Ser. 137, 451 (1999).

15. M. Imbert, Astron. Astrophys., Suppl. Ser. 140, 79 (1999).

16. S. Madsen, D. Dravins, and L. Lindegren, Astron. Astrophys.
381, 446 (2002).

17. J. R. de Medeiros and M. Mayor, Astron. Astrophys., Suppl.
Ser. 139, 433 (1999).

18. J. R. de Medeiros, S. Udry, G. Burki, et al., Astron.
Astrophys. 395, 97 (2002).

19. S. M. Rucinski, C. C. Capobianco, W. Lu, et al., Astron. J.
125, 3258 (2003).

20. E. Solano, R. Garrid, J. Fernley, et al., Astron. Astrophys.,
Suppl. Ser. 125, 321 (1997).

21. J. Storm, B. W. Carney, W. P. Gieren, et al., Astron.
Astrophys. 415, 531 (2004).

22. S. Malaroda, H. Levato, and S. Galliani, Complejo Astronomico
El Leoncito (2004).

23. S. Udry, M. Mayor, E. Maurice, et al., Precise Stellar Radial
Velocities, Ed. by L. P. Ossipkov and I. I. Nikiforov, Astron.
Soc. Pac. Conf. Ser. 185, 383 (1999).

24. F. C. Fekel, Precise Stellar Radial Velocities, Ed. by L.
P.Ossipkov and I. I. Nikiforov, Astron. Soc. Pac. Conf. Ser. 185,
378 (1999).

25. S. Udry, M. Mayor, and D. Cueloz, Precise Stellar Radial
Velocities, Ed. by L. P. Ossipkov and I. I. Nikiforov, Astron.
Soc. Pac. Conf. Ser. 185, 367 (1999).

26. R. P. Stefanik, D. W. Latham, and G. Torres, Precise Stellar
Radial Velocities, Ed. by L. P. Ossipkov and I. I. Nikiforov,
Astron. Soc. Pac. Conf. Ser. 185, 354 (1999).

27. D. L.Nidever,G . W. Marcy,R.P. Butler, et al.,Astrophys. J.,
Suppl. Ser. 141, 503 (2002).

28. W. Dehnen and J. J. Binney, Mon. Not. R. Astron. Soc. 298, 387
(1998).

29. S. V. M. Clube, Mon. Not. R. Astron. Soc. 159, 289 (1972).

30. S. V. M. Clube, Mon. Not. R. Astron. Soc. 161, 445 (1973).

31. K. F. Ogorodnikov, Dynamics of Stellar Systems (Fizmatgiz,
Moscow, 1958; Pergamon, Oxford, 1965).

32. V. V. Bobylev, Pis$'$ma Astron. Zh. 30, 861 (2004) [Astron.
Lett. 30, 785 (2004)].

33. P. P. Parenago, Kurs zvezdno@i astronomii (Course of Stellar
Astronomy) (Gosizdat, Moscow, 1954) [in Russian].

34. R. J. Trumpler and H. F.Weaver, Statistical Astronomy (Univ.
of Calif. Press, Berkeley, 1953).

35. V. V. Bobylev, Pis$'$ma Astron. Zh. 30, 185 (2004) [Atron.
Lett. 30, 159 (2004)].

36. J. Skuljan, J. B. Hearnshaw, and P. L. Cottrell, Mon. Not. R.
Astron. Soc. 308, 731 (1999).

37. R. S. De Simone, X.Wu, and S. Tremaine, Mon. Not. R. Astron.
Soc. 350, 627 (2004).

38. C. A. Murray, Vectorial Astrometry (Hilger, Bristol, 1983).

39. V. V. Bobylev, V. V. Vityazev, and G. A. Gontcharov, Vestn.
Peterb. Univ., Ser. 1, issue 4, N 25, 111 (2003).

40. R. Fux, Astron. Astrophys. 373, 511 (2001).

41. G. M$\ddot{u}$hlbauer and W. Dehnen, Astron. Astrophys. 401,
975 (2003).

42. C. Babusiaux and G. Gilmor, astro-ph/0501383 (2005).

43. L. R. Bedin, G . Piotto, I. R. King, et al., Astron. J. 126,
247 (2003).

44. M. V. Zabolotskikh, A. S. Rastorguev, and A. K. Dambis,
Pis$'$ma Astron. Zh. 28, 516 (2002) [Astron. Lett. 28, 454
(2002)].

45. D. Fern$\acute{a}$ndez, F. Figueras, and J. Torra, Astron.
Astrophys. 372, 833 (2001).

46. F. Pont, M. Mayor, and G. Burki, Astron. Astrophys. 285, 415
(1994).

47. M. R. Metzger, J. A. R. Caldwell, and P. L. Schechter, Astron.
J. 115, 635 (1998).

48. K. Rohlfs, Lectures in Density Waves (Springer, Berlin, 1977;
Mir,Moscow, 1980).

49. J. Torra, D. Fern$\acute{a}$ndez, and F. Figueras, Astron.
Astrophys. 359, 82 (2000).

50. A. M. Mel$'$nik, A. K. Dambis, and A. S. Rastorguev, Pis$'$ma
Astron. Zh. 27, 611 (2001) [Astron. Lett. 27, 521 (2001)].

51. S. P. Rybka, Kinematika Fiz. Nebesnykh Tel 20, 437 (2004).

\bigskip

Translated by A. Dambis

\clearpage \onecolumn

\newpage

\begin{table}

\def\baselinestretch{1}\normalsize\small
\centerline{{\bf Table 1. }Distribution of intrinsically bright
stars (in percent) in the Galactic octants and heliocentric
distance}
\medskip
\begin{center}
  \begin{tabular}{|c|c|c|c|c|}
  \hline
 Octant & $<$400 пк, Mv$<-2$ & $<$500 пк, Mv$<-2$ & 400-700 пк, Mv$<-2$ & $>$800 пк, Mv$<-4$ \\
  \hline
0-45    & 11.1 & 9.9 & 9.3 & 10.5 \\
45-90   & 11.0 & 12.8 & {\bf17.7} & 14.1 \\
90-135  & 12.5 & {\bf14.2} & {\bf17.4} & {\bf19.7} \\
135-180 & 11.4 & 10.9 & 10.8 & 11.2 \\
180-225 & 13.1 & {\bf13.7} & {\bf13.0} & 12.3 \\
225-270 & 11.4 & {\bf13.6} & {\bf14.8} & 8.9 \\
270-315 & 12.5 & 11.5 & 9.5 & {\bf15.1} \\
315-360 & {\bf 17.0} & 13.3 & 7.5 & 8.2 \\
  \hline
Total number of stars & 534 & 815 & 817 & 813 \\
  \hline

  \end{tabular}
  \end{center}

\end{table}
\newpage
\begin{table}[d]
\centerline{{\bf Table 2.} Format of the current release of the
OSACA catalog.}
\begin{center}
\begin{tabular}{|l|l|}

\hline

HIPPARCOS number                                         &  I6    \\
Right ascension in decimal degrees                             &  F11.6 \\
Declination in decimal degrees                                      &  F11.6 \\
Parallax in milliarcseconds                                     &  F7.2  \\
Standard error of parallax in milliarcseconds                      &  F7.2  \\
Fractional error of parallax                                  &  F6.2  \\
Decimal logarithm of parallax in milliarcseconds       &  F6.2  \\
Proper motion in right ascension in 0.001$''$/year            &  F9.1  \\
Standard error of the proper motion in right ascension in 0.001$''$/year &  F6.1  \\
Proper motion in declination in 0.001$''$/year     &  F9.1  \\
Standard error of the proper motion in declination in 0.001$''$/year          &  F6.1  \\
Radial velocity in km/s                                            &  F7.1  \\
Standard error of radial velocity in km/s                                   &  F4.1  \\
$V$ magnitude in Johnson system (from HIPPARCOS)             &  F6.2  \\
$B$ magnitude in Johnson system (from HIPPARCOS)              &  F6.2  \\
$I$ magnitude in Johnson system (from HIPPARCOS)            &  F6.2  \\
$B-V$ color index in Johnson system (from HIPPARCOS)            &  F7.3  \\
$V-I$ color index in Johnson system (from HIPPARCOS)            &  F6.2  \\
Interstellar extinction $A_v$ [6]                                   &  F5.2  \\
Absolute magnitude $M_V$                             &  F6.1  \\
Binarity parameter: 1.single star, 2.nonsingle star        &  A1    \\
Spectral type and luminosity class (from HIC [7])         &  A13   \\
Galactic longitude in decimal degrees                           &  F11.6 \\
Galactic latitude in decimal degrees                          &  F11.6 \\
Proper motion in Galactic longitude in 0.001$''$/year        &  F8.1  \\
Proper motion in Galactic latitude in 0.001$''$/year          &  F8.1  \\
Total space velocity in km/s                             &  F7.1  \\
Velocity component $U$ in Galactic coordinate system in km/s     &  F8.1  \\
Velocity component $V$ in
Galactic coordinate system in km/s      &  F8.1  \\
Velocity component $W$ in Galactic coordinate system in km/s    &  F8.1  \\
$X$ coordinate in
Galactic coordinate system in pc                &  F8.1  \\
$Y$ coordinate
in Galactic coordinate system in pc      &  F8.1  \\
$Z$coordinate in Galactic coordinate system in pc                 &  F8.1  \\
\hline
\end{tabular}
\end{center}
\end{table}

\newpage
{
\begin{table}[d]                                                
\centerline{{\bf Table 3.} Kinematic parameters of nearby
main-sequence stars. }
\begin{center}
\begin{tabular}{|l|r|r|r|r|r|r|}\hline
Parameter  & ~Single stars~ & ~Binary stars~ & ~Single stars~ & ~Binary stars~ \\
     \cline{2-5}
   & \multicolumn{2}{|c|}{ B-V$\leq0.5$} & \multicolumn{2}{|c|}{ B-V$>0.5$} \\
     \hline
 $N_\star$          & ~1720~ & ~1100~ & ~2318~ & ~1138~ \\
 $\overline r,$~pc & ~67~ & ~51~ & ~39~ & ~30~ \\
     \hline
 $u_\odot$,km/s & $10.8\pm0.4$ & $10.4\pm0.5$ & $ 9.0\pm0.6$ & $10.8\pm0.9$ \\
 $v_\odot$,km/s & $11.8\pm0.4$ & $11.8\pm0.5$ & $20.2\pm0.6$ & $21.9\pm0.9$ \\
 $w_\odot$,km/s & $ 6.9\pm0.4$ & $ 7.3\pm0.5$ & $ 8.1\pm0.6$ & $ 7.4\pm0.9$ \\
 $V_\odot$,km/s & $17.4\pm0.4$ & $17.3\pm0.5$ & $23.5\pm0.6$ & $25.5\pm0.9$ \\
 $L_\odot$,degrees & $  48\pm1$   & $  48\pm2$   & $  66\pm2$   & $  64\pm2  $ \\
 $B_\odot$,degrees & $  24\pm1$   & $  25\pm2$   & $  20\pm2$   & $  17\pm2  $ \\
 \hline
 $M_{11}$,km s$^{-1}$ kpc$^{-1}$ & $-12\pm9$ & $ 11\pm12$ & $-21\pm18$ & $103\pm29$ \\
 $M_{12}$,km s$^{-1}$ kpc$^{-1}$ & $ 52\pm7$ & $ 33\pm9 $ & $103\pm17$ & $ 27\pm25$ \\
 $M_{13}$,km s$^{-1}$ kpc$^{-1}$ & $-16\pm9$ & $ -3\pm12$ & $-22\pm17$ & $  0\pm28$ \\
 $M_{21}$,km s$^{-1}$ kpc$^{-1}$ & $ -7\pm9$ & $-25\pm12$ & $  2\pm18$ & $ 51\pm29$ \\
 $M_{22}$,km s$^{-1}$ kpc$^{-1}$ & $ 19\pm7$ & $ 10\pm9 $ & $ 20\pm17$ & $  7\pm25$ \\
 $M_{23}$,km s$^{-1}$ kpc$^{-1}$ & $  4\pm9$ & $ 28\pm12$ & $ 15\pm17$ & $-16\pm28$ \\
 $M_{31}$,km s$^{-1}$ kpc$^{-1}$ & $  2\pm9$ & $ -9\pm12$ & $ 16\pm18$ & $  1\pm29$ \\
 $M_{32}$,km s$^{-1}$ kpc$^{-1}$ & $  3\pm7$ & $  4\pm9 $ & $-28\pm17$ & $ 19\pm25$ \\
 $M_{33}$,km s$^{-1}$ kpc$^{-1}$ & $  6\pm9$ & $-10\pm12$ & $  0\pm17$ & $-15\pm28$ \\
 \hline

 $\lambda_1$,km s$^{-1}$ kpc$^{-1}$ & $ 31\pm9$ & $ 19\pm12$ & $ 57\pm18$ & $117\pm29$ \\
 $\lambda_2$,km s$^{-1}$ kpc$^{-1}$ & $-25\pm9$ & $-21\pm12$ & $-57\pm18$ & $-15\pm29$ \\
 $\lambda_3$,km s$^{-1}$ kpc$^{-1}$ & $  8\pm9$ & $ 13\pm12$ & $ -1\pm17$ & $ -7\pm28$ \\
     \hline
 $\sigma_1$,km/s  & $ 23.09\pm0.55$ & $ 22.67\pm0.67$ & $ 33.84\pm0.76$ & $ 33.82\pm1.07$ \\
 $\sigma_2$,km/s  & $ 13.21\pm0.52$ & $ 13.20\pm0.59$ & $ 22.00\pm0.52$ & $ 23.06\pm0.73$ \\
 $\sigma_3$,km/s  & $ 10.15\pm0.37$ & $ 10.21\pm0.45$ & $ 17.75\pm0.41$ & $ 17.27\pm0.56$ \\
 $l_1, ~b_1$,degrees & $ 14\pm3$,  ~$-1\pm0$ & $ 15\pm11$, ~$-2\pm2$ & $  8\pm1$, ~~$ 0\pm0$& $  9\pm9$,  ~$-2\pm2$ \\
 $l_2, ~b_2$,degrees & $104\pm2$, ~~$ 0\pm1$ & $105\pm3$, ~~$ 4\pm2$ & $ 98\pm2$, ~~$ 5\pm1$& $ 98\pm4$, ~~$ 9\pm2$ \\
 $l_3, ~b_3$,degrees & $349\pm2$,  ~$89\pm6$ & $315\pm3$,  ~$86\pm5$ & $275\pm2$,  ~$86\pm5$& $291\pm4$,  ~$81\pm5$ \\
%
 \hline
\end{tabular}
\end{center}

\end{table}
}

\newpage
\begin{figure}[t]
{\begin{center}
  \includegraphics[width= 120mm]{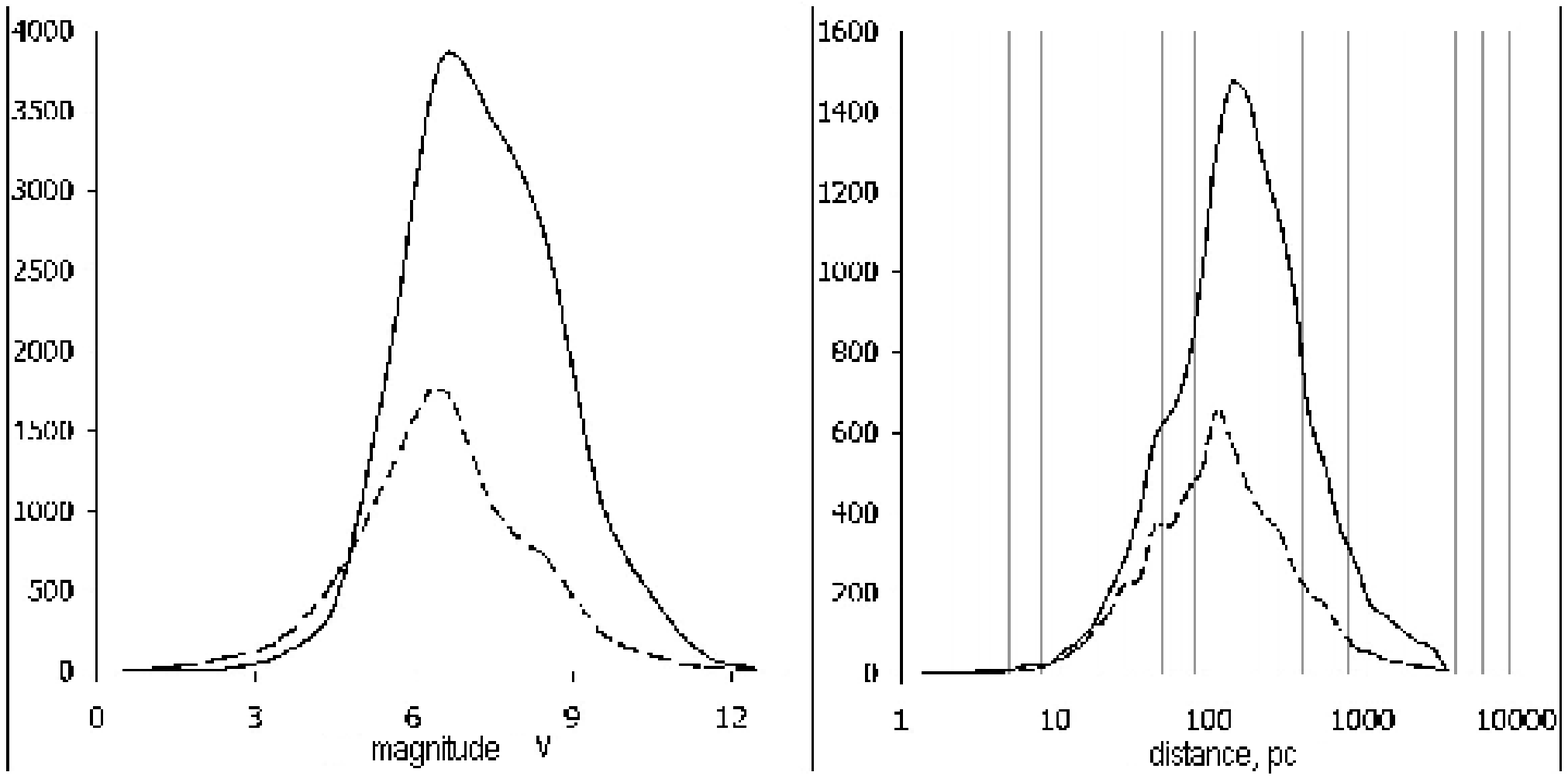}

\medskip
{\bf Fig.1.}~Distributions of OSACA stars in absolute magnitude
and heliocentric velocity, for single (solid) and multiple
(dashed) stars.
\end{center}}
\end{figure}

\begin{figure}[p]
{\begin{center}
  \includegraphics[width= 120mm]{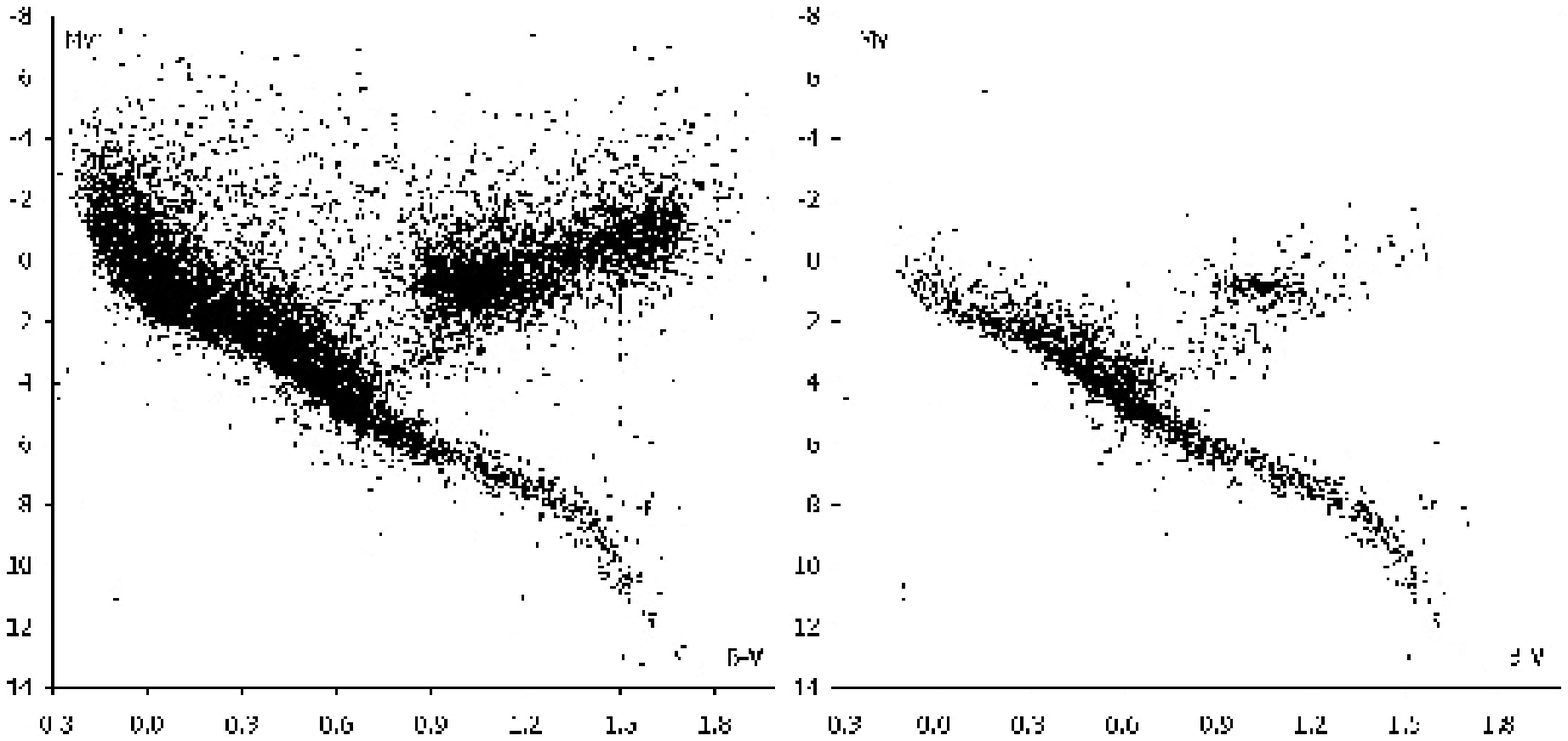}

\medskip
{\bf Fig.2.}~Distributions of OSACA stars in absolute magnitude
and heliocentric velocity, for single (solid) and multiple
(dashed) stars.
\end{center}}
\end{figure}

\begin{figure}[p]
{\begin{center}
  \includegraphics[width= 120mm]{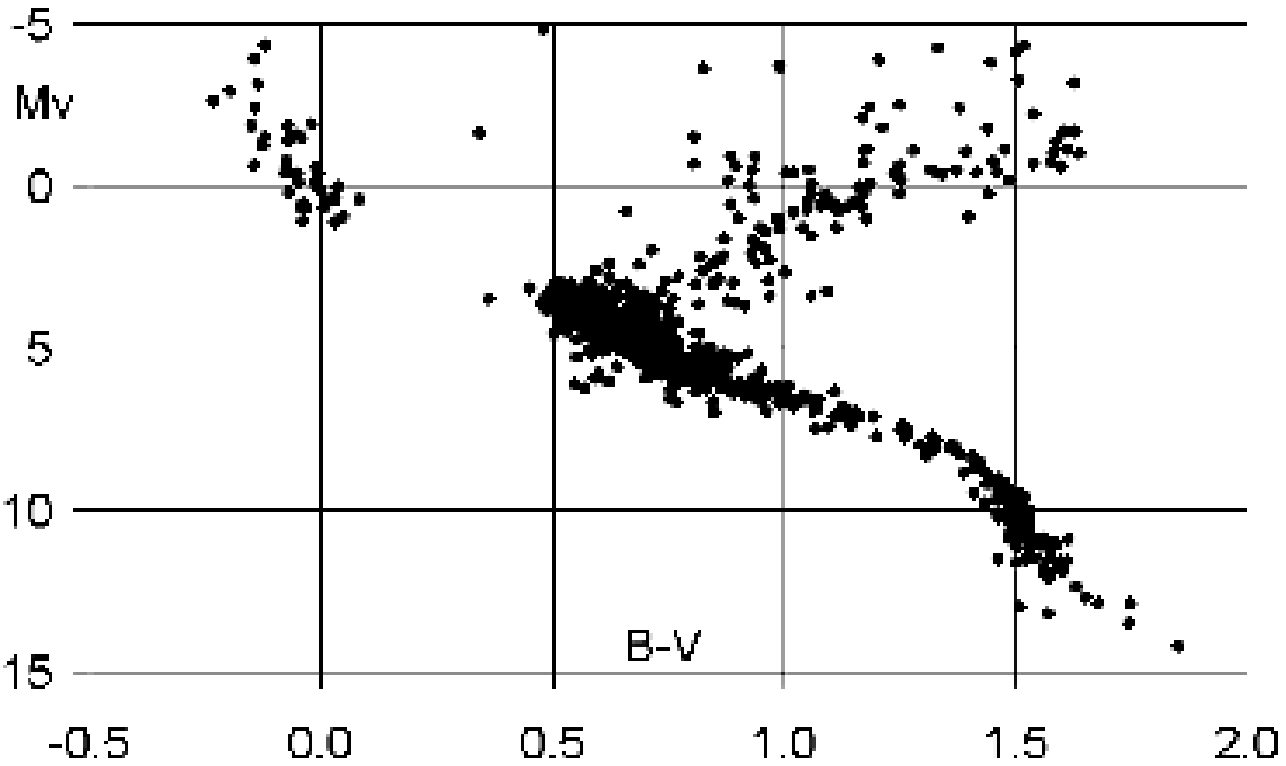}

\medskip
{\bf Fig.3.}~$(B-V)-M_V$ color index--absolute magnitude diagram
for IAU-Keck standard stars.
\end{center}}
\end{figure}

\begin{figure}[p]
{\begin{center}
  \includegraphics[width= 100mm]{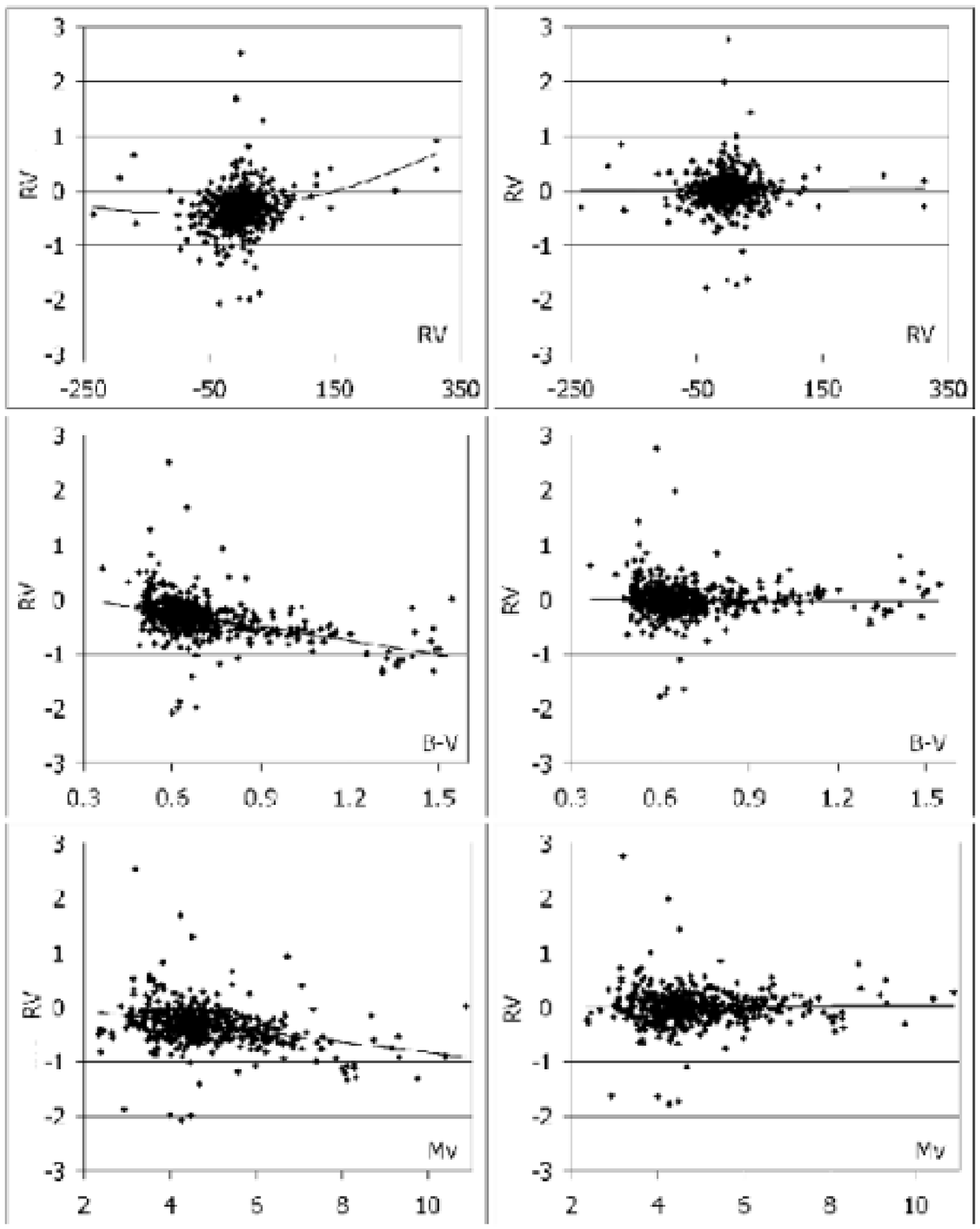}

\medskip
{\bf Fig.4.}~CORAVEL-CfA minus standard radial-velocity
differences for 484 stars in common as a function of various
parameters before (left-hand plots) and after (righthand plots)
allowance for systematic effects.
\end{center}}
\end{figure}

\begin{figure}[p]
{\begin{center}
  \includegraphics[width= 120mm]{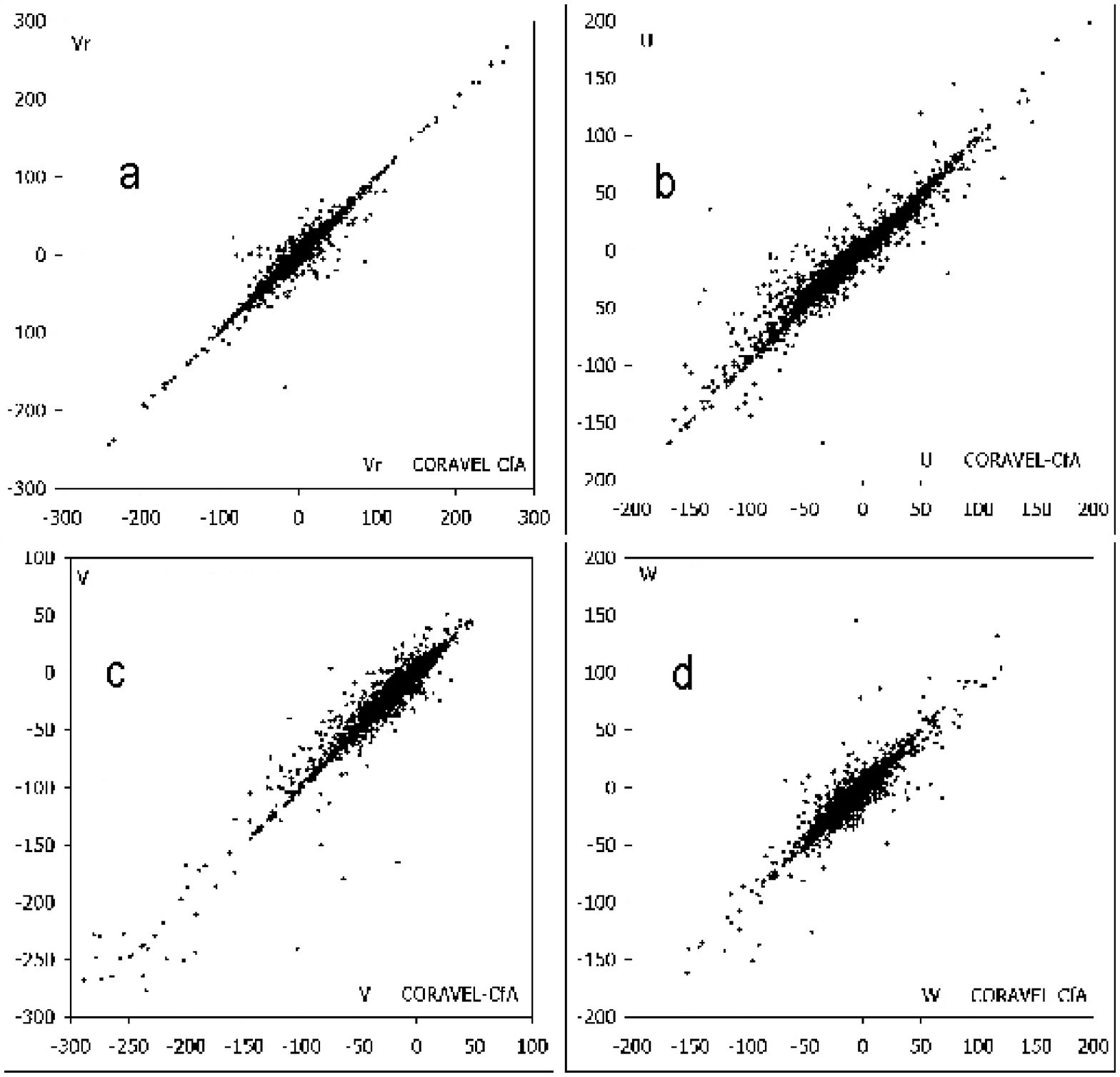}

\medskip
{\bf Fig.5.}~Comparison of the (a) radial velocities, (b) $U$
velocity components, (c) $V$ velocity components, and (d) $W$
velocity components from the CORAVEL-CfA catalog with the
corresponding velocities calculated from other radial-velocity
sources (a total of 3400 stars in common) for.
\end{center}}
\end{figure}

\begin{figure}[p]
{\begin{center}
  \includegraphics[width= 120mm]{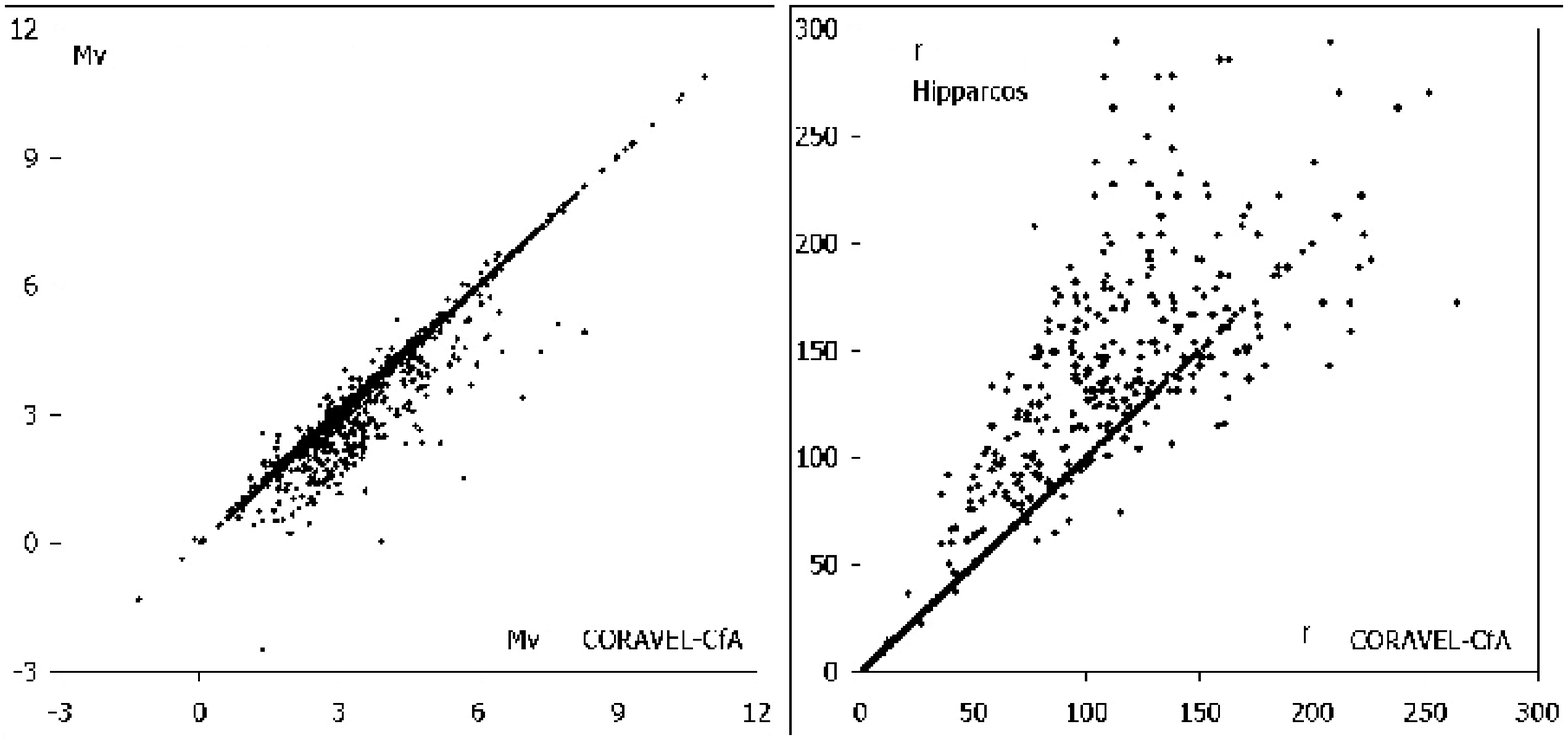}

\medskip
{\bf Fig.6.}~Left: comparison of absolute magnitudes from the
CORAVEL-CfA catalog with the absolute magnitudes calculated from
the HIPPARCOS parallaxes with allowance for interstellar
extinction (3400 stars in common). The difference is due to the
differences in the allowance for extinction and the use of
inaccurate photometric parallaxes in the CORAVEL-CfA catalog.
Right: heliocentric distances of stars of the CORAVEL-CfA catalog
compared to their heliocentric distances from the HIPPARCOS
catalog.
\end{center}}
\end{figure}

\begin{figure}[p]
{\begin{center}
  \includegraphics[width= 100mm]{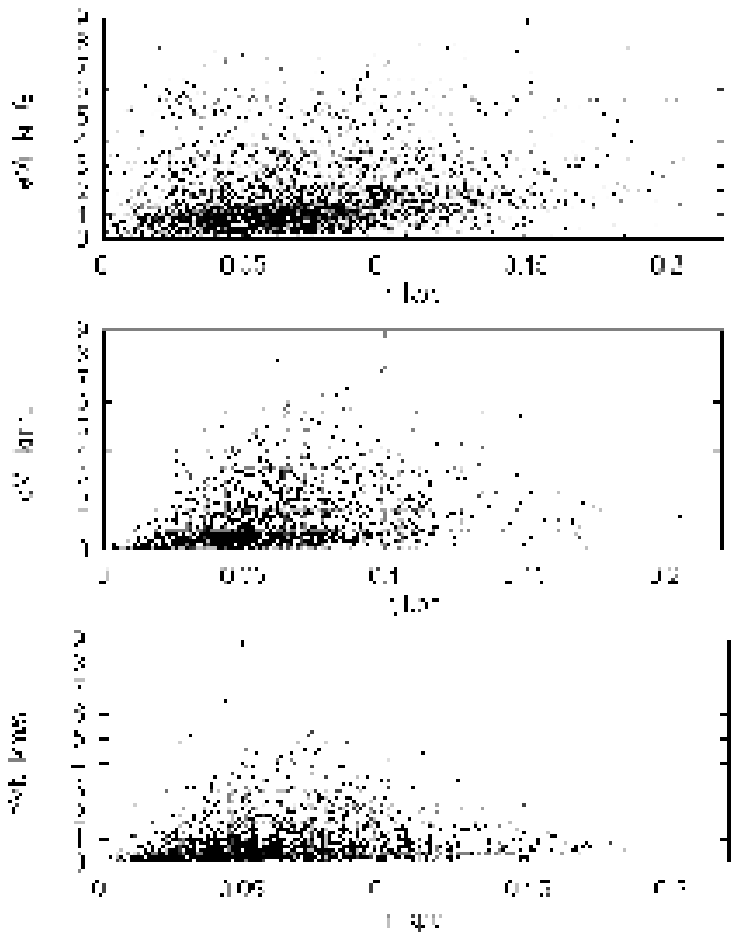}

\medskip
{\bf Fig.7.}~Random errors of the velocities of nearby
main-sequence stars as a function of heliocentric distance.
\end{center}}
\end{figure}

\begin{figure}[p]
{\begin{center}
  \includegraphics[width= 80mm]{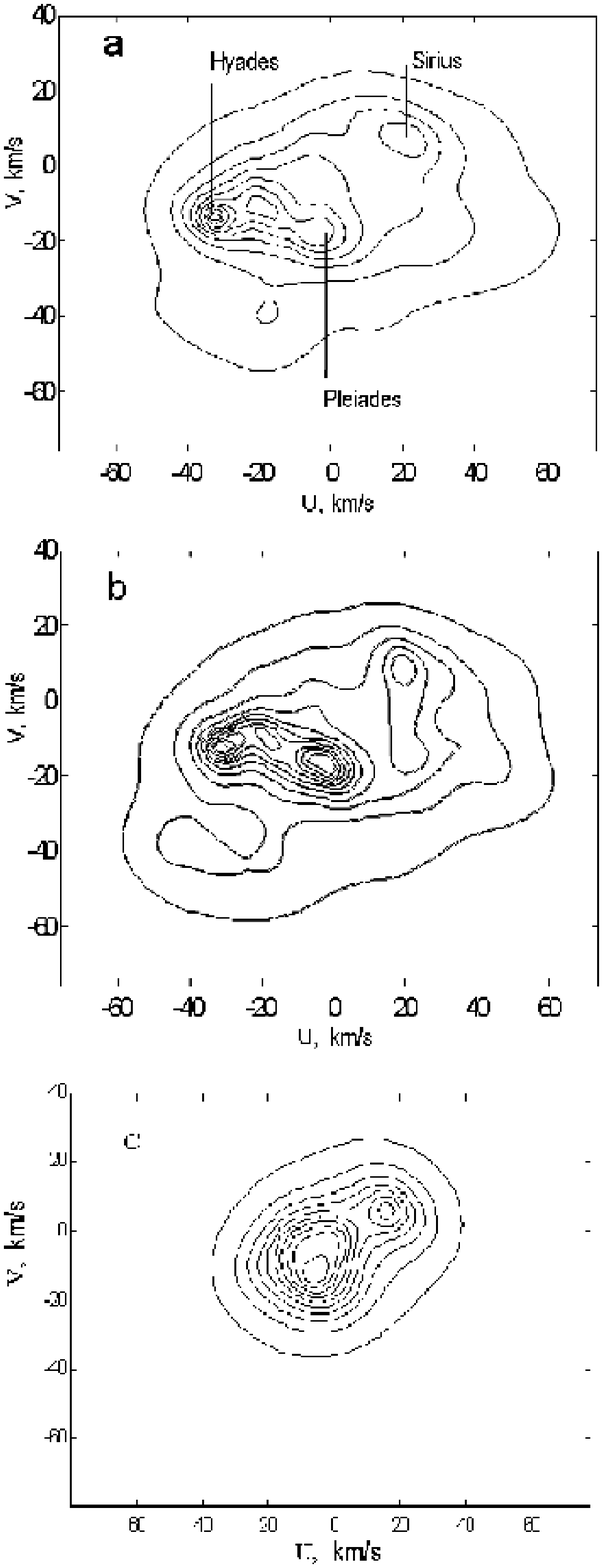}

\medskip
{\bf Fig.8.}~Distributions of $U$ and $V$ (relative to the local
standard of rest) for (a) single stars, (b) nearby main sequence
binaries with ($B-V>0.5$), and (c) stars with a negative $K$
effect.
\end{center}}
\end{figure}

\end{document}